\documentclass{article}
\usepackage{arxiv}

\usepackage{times}
\usepackage{soul}
\usepackage{url}
\usepackage[utf8]{inputenc}
\usepackage[small]{caption}
\usepackage{graphicx}
\usepackage{amsmath}
\usepackage{amsthm}
\usepackage{booktabs}
\usepackage{algorithm}
\usepackage{algorithmic}
\usepackage{bbm}
\usepackage{bm}
\usepackage{amsfonts}
\usepackage{comment}
\usepackage{algorithm}
\usepackage{algorithmic}
\usepackage{color}
\usepackage{natbib}
\usepackage{here}
\theoremstyle{definition}

\def \x {\bm{x}}
\def \positiveData {{\{\bm{x}_i\}_{i=1}^{n}}}
\def \unlabeledData {{\{\bm{x}'_i\}_{i=1}^{n'}}}
\def \Rpositive {{\mathbb{R}^{+}}}

\def \iid {\overset{\mathrm{i.i.d.}}{\sim}}

\def \Reg {\mathcal{R}}

\newcommand*{\email}[1]{\texttt{#1}}
\newcommand*{\affaddr}[1]{#1} 

\title{Identifying Different Definitions of Future\\
in the Assessment of Future Economic Conditions:\\
Application of PU Learning and Text Mining}

\author{Masahiro Kato}

\author{%
Masahiro Kato\\
\affaddr{The University of Tokyo}\\
\email{mkato@ms.k.u-tokyo.ac.jp}
}

\begin{document}

\maketitle

\begin{abstract}
The \emph{Economy Watcher Survey}, which is a market survey published by the Japanese government, contains \emph{assessments of current and future economic conditions} by people from various fields. Although this survey provides insights regarding economic policy for policymakers, a clear definition of the word ``\emph{future}'' in future economic conditions is not provided. Hence, the assessments respondents provide in the survey are simply based on their interpretations of the meaning of ``future.'' This motivated us to reveal the \emph{different interpretations of the future} in their judgments of future economic conditions by applying \emph{weakly supervised learning} and \emph{text mining}. In our research, we separate the assessments of future economic conditions into \emph{economic conditions of the near and distant future} using \emph{learning from positive and unlabeled data} (\emph{PU learning}). Because the dataset includes data from several periods, we devised new architecture to enable neural networks to conduct PU learning based on the idea of \emph{multi-task learning} to efficiently learn a classifier. Our empirical analysis confirmed that the proposed method could separate the future economic conditions, and we interpreted the classification results to obtain intuitions for policymaking. 
\end{abstract}

\section{Introduction}
The \emph{Economy Watcher Survey} is a market survey published by the Japanese government. The data consists of two types of assessments of economic conditions, \emph{current and future economic conditions}, with five ranks. Although this survey provides policymakers with deep insights, it is difficult to interpret the assessments of future economic conditions because the meaning of \emph{future} is not clearly defined and the definition thereof relies on the respondent's interpretation. Therefore, to obtain a clear understanding of survey participants’ expectations, our approach was to classify assessments of future economic conditions into those pertaining to the \emph{near and distant future}, respectively. This led us to propose a novel method that uses text data and a machine-learning algorithm in an attempt to grasp these expectations with respect to future economic conditions using data from the \emph{Economy Watcher Survey}.
For the classification task, we apply an algorithm that \emph{learns from positive and unlabeled data} (\emph{PU learning}), which is a machine-learning algorithm that enables us to train a classifier only from positive and unlabeled data.
 
Among studies of economic trends, methods using information contained in text data have become popular. Pioneering methods in this field are \citet{doi:10.1111/j.1540-6261.2007.01232.x,RePEc:bla:jfinan:v:63:y:2008:i:3:p:1437-1467}, which involved the construction of sentimental indexes from articles of a column in the \emph{Wall Street Journal} and an analysis of the predictability of the stock market. \citet{article0} predicted the residential price by using the number of searches on Google. \citet{article1} also constructed real-time inflation expectations from search queries on Google. 

PU learning is an algorithm of weakly supervised learning \citep{elkan2008learning,ward2009presence,blanchard2010semi,nguyen2011positive}. In the section describing the problem setting, we consider a situation in which only positive and unlabeled data exist, and use only these data to train a binary classifier. PU learning has two scenarios known as \emph{censoring scenario} and \emph{case-control scenario} \cite{elkan2008learning}. In the study presented in this paper, we only focus on the \emph{case-control} scenario, in which positive data are obtained separately from unlabeled data, and unlabeled data are sampled from the entire population. In this study, we construct our algorithm on the basis of subsequent research known as \emph{unbiased PU learning} \citep{ICML:duPlessis+etal:2015}, which minimizes the \emph{unbiased estimator} of the classification risk.

After classifying the assessments of future economic conditions into those relating to the near and distant future, we calculated the averaged ranks for both the near and distant future. As a result, we found that a significant difference exists between economic conditions relating to these two future periods. This result infers the possibility that people's definition of the future differs. This fact is important from the viewpoint of economics. In macroeconomics, a researcher may be interested in the possibility of controlling people's expectations of the market. Our empirical analysis reports the fact that assessments of the economic conditions of the distant future were mainly based on economic fundamentals such as the population and diplomatic relationships. 

In the following sections, we describe our problem setting and propose an algorithm that solves the problem. Subsequently, we present the results and interpretations of our empirical analysis.

\section{Problem Setting}
We consider the binary classification of text data. In the following parts, we describe the dataset and classification problem in detail.

\subsection{Economy Watchers Survey}
In our analysis, we used the \emph{Economy Watchers Survey}, a dataset that contains text data and is published by the Japanese government \footnote{Particulars of the dataset are provided on the homepage of the Japanese government, \url{https://www5.cao.go.jp/keizai3/watcher-e/index-e.html}. The survey enlists the cooperation of people holding employment positions that enable them to observe activity closely related to the regional economy. We downloaded the dataset from the page.}. The purpose of this survey is to enable the region-by-region economic trends to be grasped accurately. This survey consists of two assessments, an assessment of current and future economic conditions with the possibility of entering sentences to motivate the answers by providing reasons. Respondents evaluated the current and future economic conditions by five ranks, $0,1,2,3,4$. The evaluation $0$ means ``worse'' or ``will get worse'' compared with a previous period. The evaluation $4$ means ``better' or ``will get better'' compared with a previous period. The evaluation $2$ represents a neutral position on the assessment of economic conditions.

\paragraph{Interpretation of Assessment of Future Economic Conditions:}
Assessments of current and future economic conditions provide us with deep insights into economic reality. However, in the questionnaire, there is no clear definition of the concept of the ``future'' with respect to future economic conditions. Hence, different people interpret the duration of ``future'' in their own way. Whereas one person may imagine the future as just one week, the ``future'' might be a few months for another person. Therefore, to analyze the assessments more accurately, we need to classify assessments of future economic conditions as being either near or distant economic conditions. 

\subsection{Classification of an Assessment of Future Economic Conditions}
\begin{figure}[t]
\begin{center}
\includegraphics[width=5cm]{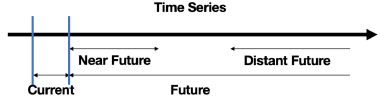}

 \caption{Our assumed definition of the time structure of assessments.}
 \label{fig:time_structure}
 \end{center}
\end{figure}

To classify future economic conditions into those expected to occur in either the near or distant future, we propose assuming that current economic conditions share similar sentences with those expected in the near future. Our classification strategy is to regard current economic conditions as positive data and future economic conditions as unlabeled data, which potentially consists of positive and negative data. In this paper, positive data are assessments of the current economic conditions and those expected in the near future, whereas negative data are assessments of economic conditions foreseen to prevail in the distant future. We illustrate the relationship between assessments of current and future economic conditions of our assumption on Figure~\ref{fig:time_structure}. We train our classifier only from positive and unlabeled data by using an algorithm that employs PU learning. Therefore, the goal of this problem is to classify $\bm{x} \in \mathcal{X} \subset \mathbb{R}^d$ into one of the two classes $\{-1, +1\}$, where $+1$ denotes assessments of current economic conditions and those expected in the near future (positive data) and $-1$ denotes economic conditions relating to the distant future (negative data). 

\subsection{Data Generating Process of Economy Watchers Survey}
Let us describe the data generating process of our problem. Let us assume that we have $n$ data points at $t$-th period and denote the $i$-th text data as $\bm{x}_i \in \mathcal{X} \subset \mathbb{R}^d$. If the target of text data $\bm{x}_i$ describes current or near future economic conditions, we attach a positive label, i.e., $y_i=+1$. If the target of text data $\bm{x}_i$ describes distant future economic conditions, we attach a negative label, i.e., $y_i=-1$. However, in the dataset, we can only observe positive data, and unlabeled data, which includes both positive and negative data. In addition, if the text data $\bm{x}_i$ belongs to a period $t\in\{1,...,T\}$, we denote the fact as $z_i=t$. Using these notations, we define our data generating process as follows:
\begin{align*}
&\positiveData\iid p(\bm{x}|y=+1, z=t),\ \unlabeledData\iid p(\bm{x}|z=t),
\end{align*}
where $\positiveData$ and $\unlabeledData$ denote the positive and negative data at $t$-th period, and $p(\bm{x}|z=t)$ can be decomposed as 
\begin{align*}
p(\bm{x}|z=t)=&p(y=+1|z=t)p(\bm{x}|y=+1, z=t)\\
& + p(y=-1|z=t)p(\bm{x}|y=-1, z=t).
\end{align*}

\section{Learning from Positive and Unlabeled Data with Time Series Data}
To classify data consisting only of positive and unlabeled data, we propose using \emph{multi-task PU learning} (\emph{MTPU}). In this section, we provide details of the proposed algorithm.

\subsection{Learning from Positive and Unlabeled Data}
Before explaining our model, let us explain the standard setting of PU learning. In PU learning, we consider a binary classification problem to classify $\bm{x} \in \mathcal{X} \subset \mathbb{R}^d$ into one of the two classes $\{-1, +1\}$. We assume that there exists a joint distribution $p(\x, y)$, where $y \in \{-1, +1\}$ is the class label of $\x$. PU learning relies on two distinct sampling schemes, namely the censoring scenario and case-control scenario \citep{elkan2008learning}. The PU learning framework we use in this study is the case-control scenario, in which we suppose access to a positive dataset $\positiveData\iid p(\bm{x}|y=+1)$ and an unlabeled dataset $\unlabeledData\iid p(\bm{x})$. Let $\ell: \mathbb{R}\times \{\pm1\}\to \Rpositive$ be a loss function, where $\Rpositive$ is the set of non-negative real values, and $\mathcal{F}$ be the set of measurable functions from $\mathcal{X}$ to $[\epsilon, 1-\epsilon]$, where $\epsilon \in (0, 1/2)$ is a small positive value. This constant $\epsilon$ is introduced to ensure the following optimization problem is well-defined based on the result of \cite{kato2018learning}. Here, \citet{ICML:duPlessis+etal:2015} showed that the classification risk of $f \in \mathcal{F}$ can be expressed as
\begin{align}
\label{true}
R_{\mathrm{PU}}&(f)\nonumber=p(y=+1) \mathbb{E}_\mathrm{p}[\ell(f(X), +1)]\nonumber\\
& -p(y=-1)\mathbb{E}_\mathrm{p}[\ell(f(X), -1)] +\mathbb{E}_\mathrm{u}[\ell(f(X), -1)],
\end{align}
where $\mathbb{E}_\mathrm{p}$ and $\mathbb{E}_\mathrm{u}$ are the expectations over $p(\bm{x}|y=+1)$ and $p(\bm{x})$, respectively. The above formulation of PU learning provides the unbiased risk of the classification problem.

\subsection{Multi-Task Non-negative PU learning for Time Series Data}
\begin{figure}[t]
\begin{center}
   \includegraphics[width=70mm]{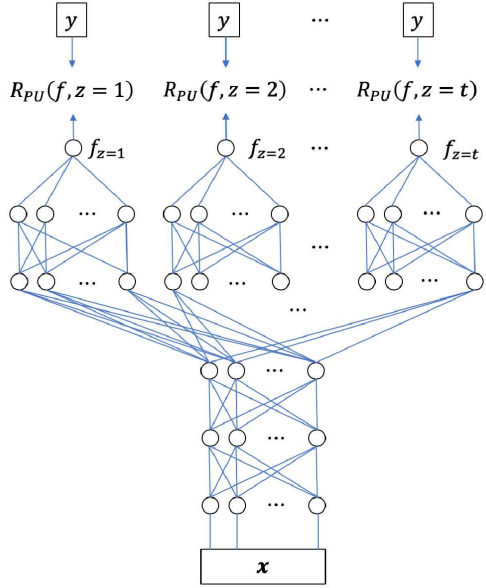}
 \caption{Neural network model for multi-task learning for PU learning. The models share one shared network with $3$ layers.}
 \label{fig:network_structure}
 \end{center}
\end{figure}

In addition to the standard setting of PU learning, we could take the time structure into account. The Economy Watcher Survey comprises monthly data, with approximately $2,600$ records for each month. Here, we would need to use different classifiers for the data included in each month for the following two reasons. First, the model can vary across periods. Second, we would not be able to include data of the $(t+1)$-th period to train a model of data of the $t$-th period because the data of the $(t+1)$-th period might have information of the data of the $t$-th period. This made it necessary to use different models across different periods. For $z=t$, we denote the model as $f_{z=t}$ and the risk as follows:
\begin{align*}
&R_{\mathrm{PU}}(f_{z=t}, z=t)=p(y=+1|z=t) \mathbb{E}_\mathrm{p, t}[\ell(f(X), +1)]\nonumber\\
& -p(y=-1|z=t)\mathbb{E}_\mathrm{p, t}[\ell(f(X), -1)] +\mathbb{E}_\mathrm{u, t}[\ell(f(X), -1)],\nonumber,
\end{align*}
where $\hat{\mathbb{E}}_{\mathrm{p, t}}$ denotes the averaging operator over positive data, $\hat{\mathbb{E}}_{\mathrm{u, t}}$ denotes averaging over the unlabeled data at the $t$-th period. We additionally introduce a model for \emph{multi-task learning} to PU learning. Multi-task learning is proposed to train neural networks efficiently by using the common features across different tasks \cite{Caruana1997}. If a common feature exists across periods, we can train our models more efficiently by sharing the common feature among models $f_{z=t}$ for $t=1,...,T$ through the layers named \emph{shared layers}, the structure of which is shown in Figure \ref{fig:network_structure}. 
We named this model \emph{MTPU}. Details of its structure are provided in the section for empirical experiments. 

\begin{table*}[t]
\caption{Averaged assessments for each period and each type of economic condition. For averaged assessments of near and distant economic conditions, we conducted a two-sample t-test. A significant difference between the mean values of the assessment is indicated by superscript $*$ in the table. One $*$ means that the null hypothesis of the two-sample t-test is rejected at the $5\%$ significance level, whereas two $*$s means that the null hypothesis of the two-sample t-test is rejected at the $1\%$ significance level.}
\label{table}
\begin{center}
\scalebox{0.65}[0.65]{
  \begin{tabular}{ll||ll|ll|ll|ll}
      \hline
          &  &MTPU  & &Original & &PU1 &  &PU2 &  \\
    \hline
          &  & NF & DF &  Current & Future& NF & DF & NF & DF \\
    \hline
    Jan. & 2016 & 1.842** &1.925** &1.996 &1.864 &1.889 &1.857&                                 1.846 &                                    1.964\\
    Feb. & 2016 & 1.701** &1.720** &1.949 &1.780 &1.749 &1.756&                                 1.733 &                                    1.876 \\
    Mar. & 2016 & 1.768* &1.694* &1.870 &1.800 &1.805 &1.731&                                 1.691 &                                    1.804 \\
    Apr. & 2016 & 1.606** &1.668** &1.822 &1.719 &1.764 &1.625&                                 1.717 &                                    1.684 \\
    May & 2016 & 1.574** &1.736** &1.896 &1.683 &1.661 &1.680&                                 1.582 &                                    1.808 \\
    June & 2016 & 1.588 &1.622 &1.656 &1.632 &1.671 &1.531&                                 1.588 &                                    1.685 \\
    July & 2016 & 1.721** &1.784** &1.887 &1.804 &1.797 &1.864&                                 1.725 &                                    1.860  \\
    Aug. & 2016 & 1.770** &1.906** &1.904 &1.815 &1.789 &1.863&                                 1.695 &                                    1.887 \\
    Sept. & 2016 & 1.661** &1.860** &1.961 &1.781 &1.700 &1.767&                                 1.650 &                                    1.887 \\
    Oct. & 2016 & 1.789** &1.914** &1.985 &1.852 &1.785 &1.878&                                 1.762 &                                    1.906 \\
    Nov. & 2016 & 2.028 &1.865 &1.978 &1.960 &1.936 &1.956&                                 1.944 &                                    1.948 \\
    Dec. & 2016 & 2.139** &1.975** &1.974 &2.095 &2.053 &2.074&                                 2.151 &                                    1.996 \\
    \hline
    Jan. & 2017 & 2.036 &1.888 &1.997 &1.959 &1.897 &1.932&                                 2.008 &                                    2.008 \\
    Feb. & 2017 & 1.947** &1.992** &2.091 &1.969 &1.886 &2.024&                                 1.886 &                                    2.033  \\
    Mar. & 2017 & 2.181* &1.881* &1.967 &2.040 &2.122 &1.984&                                 2.157 &                                    1.968 \\
    Apr. & 2017 & 2.176 &1.963 &2.034 &2.039 &2.049 &1.959&                                 2.135 &                                    1.955 \\
    May & 2017 & 2.077* &1.963* &2.080 &2.013 &1.988 &2.041&                                 2.061 &                                    2.008 \\
    June & 2017 & 2.086* &1.939* &2.084 &2.022 &1.951 &1.971&                                 2.016 &                                    2.078 \\
    July & 2017 & 2.162 &1.980 &2.034 &2.055 &2.077 &2.033&                                 2.016 &                                    1.972 \\
    Aug. & 2017 & 2.024 &1.904 &2.011 &1.975 &1.996 &1.952&                                 2.008 &                                    1.956 \\
    Sept. & 2017 & 2.041 &1.914 &2.032 &2.005 &1.943 &1.967&                                 1.931 &                                    2.033 \\
    Oct. & 2017 & 1.909** &2.040** &2.175 &1.998 &1.822 &2.048&                                 1.905 &                                    2.139 \\
    Nov. & 2017 & 2.278 &2.045 &2.076 &2.125 &2.173 &2.020&                                 2.121 &                                    2.061 \\
    Dec. & 2017 & 2.140** &2.137** &2.067 &2.171 &2.108 &2.068&                                 2.240 &                                    2.177 \\
    \hline
    Jan. & 2018 & 1.935** &1.951** &2.124 &1.957 &1.874 &2.016&                                 1.935 &                                    2.029\\
    Feb. & 2018 & 1.831** &1.938** &2.125 &1.936 &1.778 &1.942&                                 1.926 &                                    1.950\\
    Mar. & 2018 & 2.052* &2.032* &2.005 &2.073 &2.105 &1.943&                                 2.073 &                                    1.992\\
    Apr. & 2018 & 2.184 &1.922 &2.071 &2.069 &2.123 &1.984&                                 2.115 &                                    2.057\\
    May & 2018 & 1.907** &1.837** &2.039 &1.900 &1.870 &1.894&                                 1.878 &                                    1.963 \\
    June & 2018 & 1.860** &1.963** &2.034 &1.910 &1.848 &1.942&                                 1.835 &                                    2.021\\
    July & 2018 & 1.883** &1.838** &1.965 &1.874  &1.785 &1.887&                                 1.895 &                                    1.919\\
    Aug. & 2018 & 1.900** &1.988** &2.034 &1.913 &1.799 &1.931&                                 1.956 &                                    2.016\\
    Sept. & 2018 & 1.735** &1.881** &2.041 &1.882 &1.861 &1.918&                                 1.682 &                                    2.000 \\
    Oct. & 2018 & 1.964** &1.834** &2.006 &1.916 &1.948 &1.785&                                 1.911 &                                    1.887\\
    Nov. & 2018 & 2.049 &1.909 &2.033 &1.980 &1.943 &1.930&                                 1.947 &                                    1.979\\
    Dec. & 2018 & 2.017* &1.862* &1.881 &1.951 &1.996 &1.900&                                 2.017 &                                    1.912\\
    \hline
    Jan. & 2016 & 1.780** &1.784** &2.016 &1.800 &1.748 &1.833&                                 1.756 &                                    1.833\\
    Feb. & 2019 & 1.871** &1.925** &2.014 &1.863 &1.917 &1.837&                                 1.829 &                                    1.95\\
    Mar. & 2019 & 1.975* &1.856* &1.943 &1.879 &2.004 &1.797&                                 1.895 &                                    1.856\\
    Apr. & 2019 & 1.852 &1.894 &1.972 &1.916 &1.877 &1.962&                                 1.797 &                                    1.928\\
    May & 2019 & 1.845** &1.762** &1.863 &1.766 &1.853 &1.713&                                 1.784 &                                    1.758\\
    June & 2019 & 1.736** &1.676** &1.868 &1.719 &1.762 &1.718&                                 1.715 &                                    1.748 \\
  \end{tabular}
  }
\end{center}
\end{table*}

\begin{figure*}[t]
\begin{center}
   \includegraphics[width=180mm]{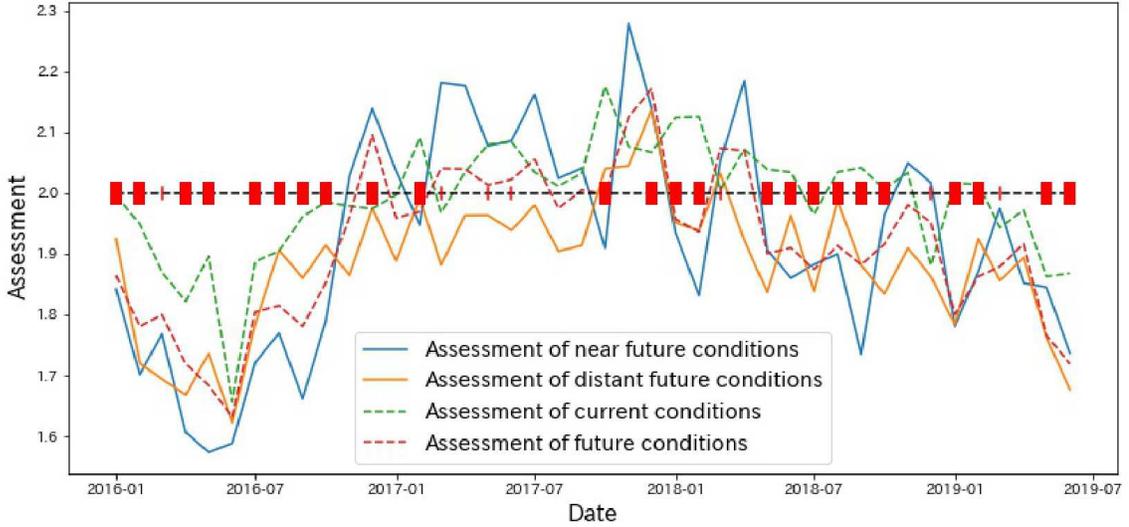}
 \caption{Plotted assessments of the economic conditions of the near and distant future and those pertaining to the present and future. The horizontal line at $y=2$ is the neutral state. The red vertical lines on the horizontal line represent the results of the two-sample t-test. The thin and bold red vertical lines represent the $5\%$ and $1\%$ significance levels, respectively.}
 \label{fig:timeseries}
 \end{center}
 \end{figure*}
\begin{figure*}[t]
\begin{center}
   \includegraphics[width=145mm]{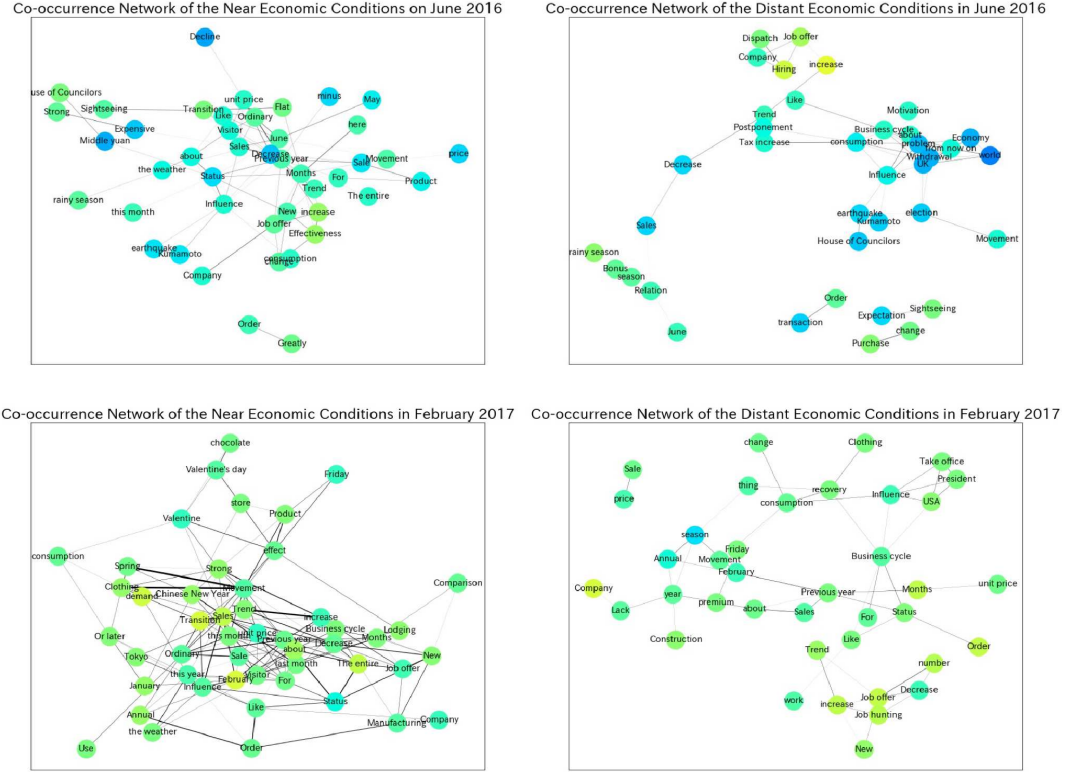}
 \end{center}
 \caption{Co-occurrence network of near and distant future economic conditions in June 2016 and February 2017. The lengths of the edges represent the value of the Jaccard coefficients. Shorter edges indicate a stronger relationship (the value of the Jaccard coefficients is larger) between the two words. The widths of the edges also represent the value of the Jaccard coefficients between the two words. The bold edges similarly signify a stronger relationship (the value of the Jaccard coefficients is larger) between the two words. The color of the nodes relates to the assessment. The yellow-green color denotes that the averaged value is $2$, i.e., the assessment is neutral. The warmer and cooler colors represent positive and negative assessments, respectively.}
 \label{fig:network}
\end{figure*}
\subsection{Sample Approximation of the Unbiased Risk}
When we train a classifier, we can naively replace the expectations with the corresponding sample averages. However, \citet{NIPS2017_6765} pointed out that the basic form of the unbiased PU learning is ineffective with a deep neural network because of over-fitting caused by the fact that the risk is not lower bounded. To implement PU learning with deep neural networks, we applied the non-negative risk proposed by \citet{NIPS2017_6765} to the empirical risk defined in (\ref{eq:emp_risk}). For a hypothesis set $\mathcal{H}$, let us define the following risk minimization problem,
\begin{align}
\label{eq:emp_risk}
&\hat{f}_{z=t}  = \mathrm{argmin}_{f_{z=t}\in\mathcal{H}}\left[\widehat{R}_{\mathrm{PU}}(f_{z=t}, z=t)+\Reg(f)\right],
\end{align}
where $\widehat{R}_{\mathrm{nnPU}}(f_{z=t}, z=t)$ is a sample approximation of $R_{\mathrm{PU}}(f_{z=t}, z=t)$ with non-negative transformation proposed by \citet{NIPS2017_6765} and $\Reg$ is a regularization term. 

\subsection{Class Prior and Selection Bias}
The remaining problem is to make a decision regarding the class prior $p(y=+1|z=t)$. The class prior $p(y=+1|z=t)$ would be different across periods $t$. Although several algorithms have been proposed to estimate the class prior \citep{IEICE:duPlessis+Sugiyama:2014,pmlr-v48-ramaswamy16,jain2016nonparametric}, the estimation is still known to be a difficult task. However, we can avoid the problematic estimation in the case of the particular goal we hope to reach. In our experiments, we assume that the class prior is $p(y=+1|z=t)=0.2$ for all periods, $t=1,2...,T$. This assumption is not realistic because the probability would have different values across the periods. However, \cite{1809.05710,kato2018learning} showed that the function $f_{z=t}$ is simply linear-proportional to the class prior, i.e., the following relationship holds even if we miss-specify the class prior:
\begin{align}
\label{eq:invariance_of_order}
&p(y=+1|x, z=t) \leq p(y=+1|x, z=t)\nonumber\\
&\Leftrightarrow f_{z=t}(x) \leq f_{z=t}(x).
\end{align} 
Therefore, even when we cannot obtain the exact value of $p(y=+1|\bm{x}, z=t)$, we can still identify the order of $p(y=+1|\bm{x}, z=t)$ with regard to $\bm{x}$. Our empirical analysis separates the assessment of future economic conditions into near and distant future economic conditions based on this property. We classify $1/5$ of data from the highest value of $f_{z=t}$ into assessments of near future economic conditions, and $1/5$ of data from the lowest value of $f_{z=t}$ into assessments of distant future economic conditions. In addition to the robustness to the miss-specified class prior, the function $f_{z=t}$ also holds the relationship \ref{eq:invariance_of_order} under the selection bias of positive data \citep{kato2018learning} if our assumption is mild. Thus, our results can reduce the influence of the miss-specified class prior and selection bias.

\section{Empirical Analysis}
In this section, we report the results of the empirical analysis of data from the Economy Watcher Survey. The survey was conducted every month starting in 2000. Our analysis only used data from January 2016 to June 2019, i.e., $42$ months' data. Each month includes approximately $2,600$ samples. The reason for the heterogeneity among the data is the lack of text in respondents’ answers. In total, we had 111,501 samples. 

We used Bag-of-Words to represent the documents as $16,914$-dimensional vectors. After vectorizing the text data, we applied PU learning with the aforementioned MTPU. In addition to the model, we also used the standard model of PU learning to compare the performance. We used this model of PU learning in two ways. First, we used all samples to train one model. Second, we prepared one model for each month. Details of the neural networks are provided in the following section. After training our classifier, we classified the assessment of future economic conditions using unlabeled data that we used for training.

\paragraph{Neural network model:} First, we describe the model we used for MTPU. The model for the shared network was a $3$-layer \emph{multilayer perceptron} (MLP) with ReLU \cite{Nair:2010:RLU:3104322.3104425} (more specifically, $16914-500-500-500$). The neural network model following the shared network was a $2$-layer MLP (more specifically, $500-500-1$) with ReLU. Next, we describe the model we used for non-negative PU learning. The model for the neural network was a $5$-layer MLP (more specifically, $16914-500-500-500-500-1$) with ReLU. We set $p(y=+1|z=t)=0.2$ for all $t\in\{1,2,...,42\}$. For both methods, we use logistic loss for the loss function $\ell$.

\subsection{Difference among the Assessments}
In this section, we report the extent to which assessments differ across current, future, near future, and distant future economic conditions. 

\paragraph{Averaged Assessments and t-test:} We report the averaged assessments of economic conditions in the near and distant future in comparison with those of the current and future. Assessments of the near and distant future economic conditions are estimated by MTPU and non-negative PU learning with neural networks. For non-negative PU learning, we used two models. The first (named PU1) entailed training one model for all samples. The second (named PU2) involved using different models for the data of different months. The results are presented in Table~\ref{table}. For each period, we show the results of the two-sample t-test with unequal variances between the assessments of the economic conditions of the near and distant future. Values for which the difference between the mean of the assessments is significant are indicated by superscript $*$ in the table. One $*$ and two $*$s mean that the null hypothesis of the two-sample t-test is rejected at the $5\%$ and $1\%$ significance levels, respectively. 

\paragraph{Visualization as a Time Series:} To facilitate a more intuitive understanding of the reported results, we plotted the averaged assessments in the time series in Figure~\ref{fig:timeseries}, where the $x$-axis corresponds to the time series, and the $y$-axis corresponds to the value of the assessment. The blue, orange, green, and red lines correspond to assessments of the economic conditions in the near future, distant future, at the present time, and in the future. The horizontal black dashed line at $y=2.0$ represents the neutral condition in the $5$-step evaluations for the economic conditions from $0$ (bad) to $4$ (good). The vertical red lines perpendicular to the line $y=2.0$ indicate that the difference between the average assessments of the economic conditions in the near and distant future is significant in the two-sample t-test. The bold vertical lines represent that the null hypothesis of the two-sample t-test is rejected at the $1\%$ significance level and the other red lines represent that two $*$ means that the null hypothesis of the two-sample t-test is rejected at the $5\%$ significance level. For example, the assessments of the economic conditions in the near future in 2017 are significantly higher than those of the distant economic conditions.

\subsection{Text Mining}
This section presents our analysis of the text based on assessments of the text data. For text mining, we use \emph{tf-idf} and the \emph{Jaccard coefficient}, which are standard techniques of natural language processing. First, we separate the assessments of the economic conditions in the near and distant future for the month in which the assessments were published, i.e., we form groups of monthly assessments. Then, we denote a set of the group of assessments as $\mathcal{M}$, and we apply tf-idf to identify the word that characterizes the document. Then, for the $50$ words with the highest tf-idf, we measure the Jaccard coefficient \cite{Manning:1999:FSN:311445}, which measures the similarity between two sets. Let $\mathcal{M}_w\in\mathcal{M}$ be a set of sentences including the word $w$. The Jaccard coefficient $J(\mathcal{M}_a, \mathcal{M}_b)$ for a word $a$ and a word $b$ can be expressed as follows:
\begin{align}
J(\mathcal{M}_a, \mathcal{M}_b) = \frac{|\mathcal{M}_a\cap \mathcal{M}_b|}{|\mathcal{M}_a\cup \mathcal{M}_b|}. 
\end{align}
Based on these results, we plotted the co-occurrence networks in Figure~\ref{fig:network}\footnote{We translated from Japanese to English using an API provided by Google (\url{https://pypi.org/project/googletrans/}). }. Because of the limitation placed on the length of the paper, we only show the network of assessments in June 2016 and February 2017. June 2016 is one of the periods in which the value of assessments greatly changed. Throughout 2017, the economic conditions of the near future are less than those in the distant future, and February 2016 is one of these periods. Because of the small size of our graphs, we placed enlarged versions of these graphs in the appendix in both English and Japanese. 

\subsection{Interpretations}
Figure~\ref{fig:network} displays words related to economic fundamentals, such as the structure of the labor supply and international politics. In other words, these results can be interpreted as meaning that assessments of the economic conditions of the near future represent the economic cycle, whereas assessments of the economic conditions of the distant future represent the economic trend. For example, the words ``U.K.'' and ``withdrawal'' appear, both of which are related to Brexit among the economic conditions of the distant future, ``Business cycle,'' and ``Trend'' in June 2016. The words ``US'' and ``President'' appear in Feb. 2017. On the other hand, the economic conditions of the near future in June 2016 and Feb, 2017 are represented by words that have less relationship with economic fundamentals such as ``rainy season'' and ``Valentin's day.'' For policymakers, this is an insightful finding because the result infers that they cannot easily change people's expectations based on economic fundamentals. 

\section{Conclusion}
In this paper, we proposed a new application of PU learning and text mining to data consisting of financial text. We developed a new model named MTPU to train neural networks efficiently using data with a time structure. Our empirical analysis showed the classification result and interpretations based on text mining and economics. The result is insightful to policymakers because the result infers that people might have a different interpretation of the definition of the future and may assess the future economic outlook differently based on their interpretations of the future.  Besides, we also found that there are different main reasons between near and distant future economic assessments.
\bibliographystyle{named}
\bibliography{arxiv.bbl}

\clearpage
\onecolumn
\appendix
\section{Graphs of Co-occurrence Network}
Because of the length limitation of the paper, we only included reduced-size versions of the graphs in the main body. Here, we present enlarged versions of Figure~\ref{fig:network}. The lengths of the edges represent the values of the Jaccard coefficients with shorter edges indicating a stronger relationship between two words (the Jaccard coefficients have larger values). The widths of the edges also represent the value of the Jaccard coefficients between two words. The bold edges similarly indicate a stronger relationship (larger values of the Jaccard coefficients) between the two words. The color of the nodes represents the averaged value of the assessments of the text in which the word appeared. Yellow-green denotes that the averaged value is $2$, i.e., the assessment is neutral. Warmer and cooler colors represent positive and negative assessments, respectively.
\clearpage

\subsection{Co-occurrence Network of Near Future Economic Conditions in June 2017}
\begin{figure}[H]
\begin{center}
\includegraphics[width=143mm]{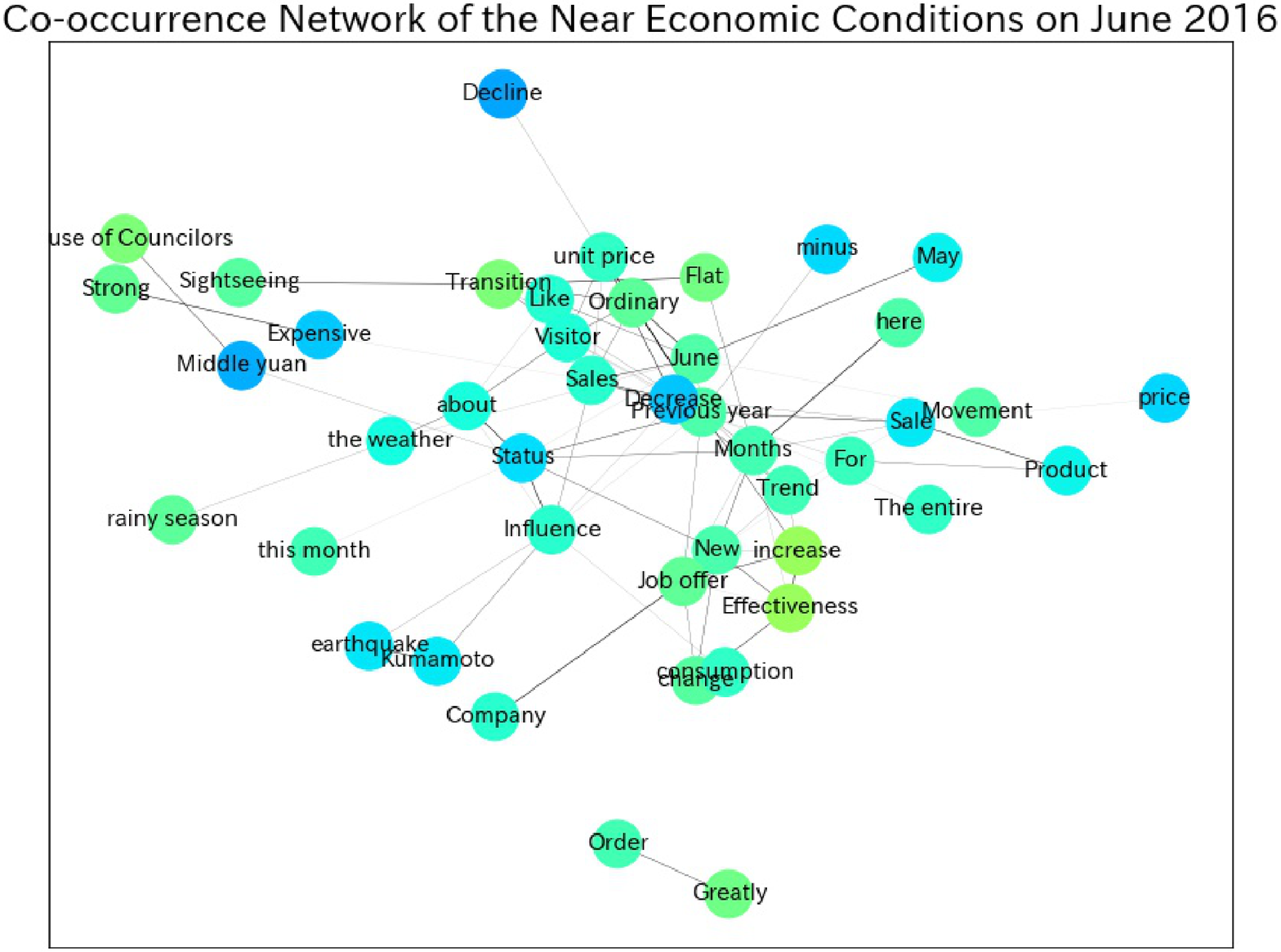}
\includegraphics[width=115mm]{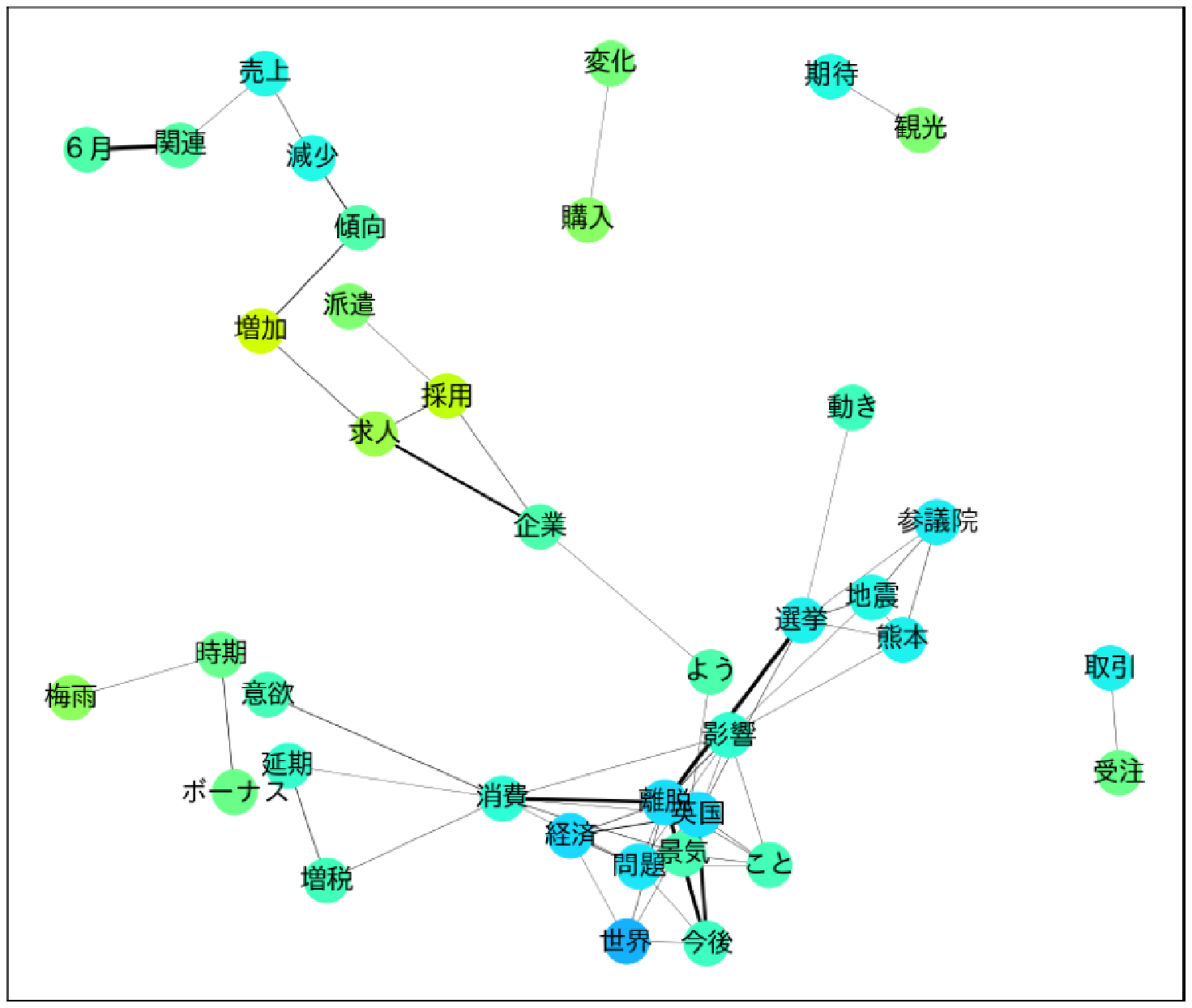}
 \end{center}
\end{figure}
\clearpage

\subsection{Co-occurrence Network of Distant Future Economic Conditions in June 2017}
\begin{figure}[H]
\begin{center}
\includegraphics[width=143mm]{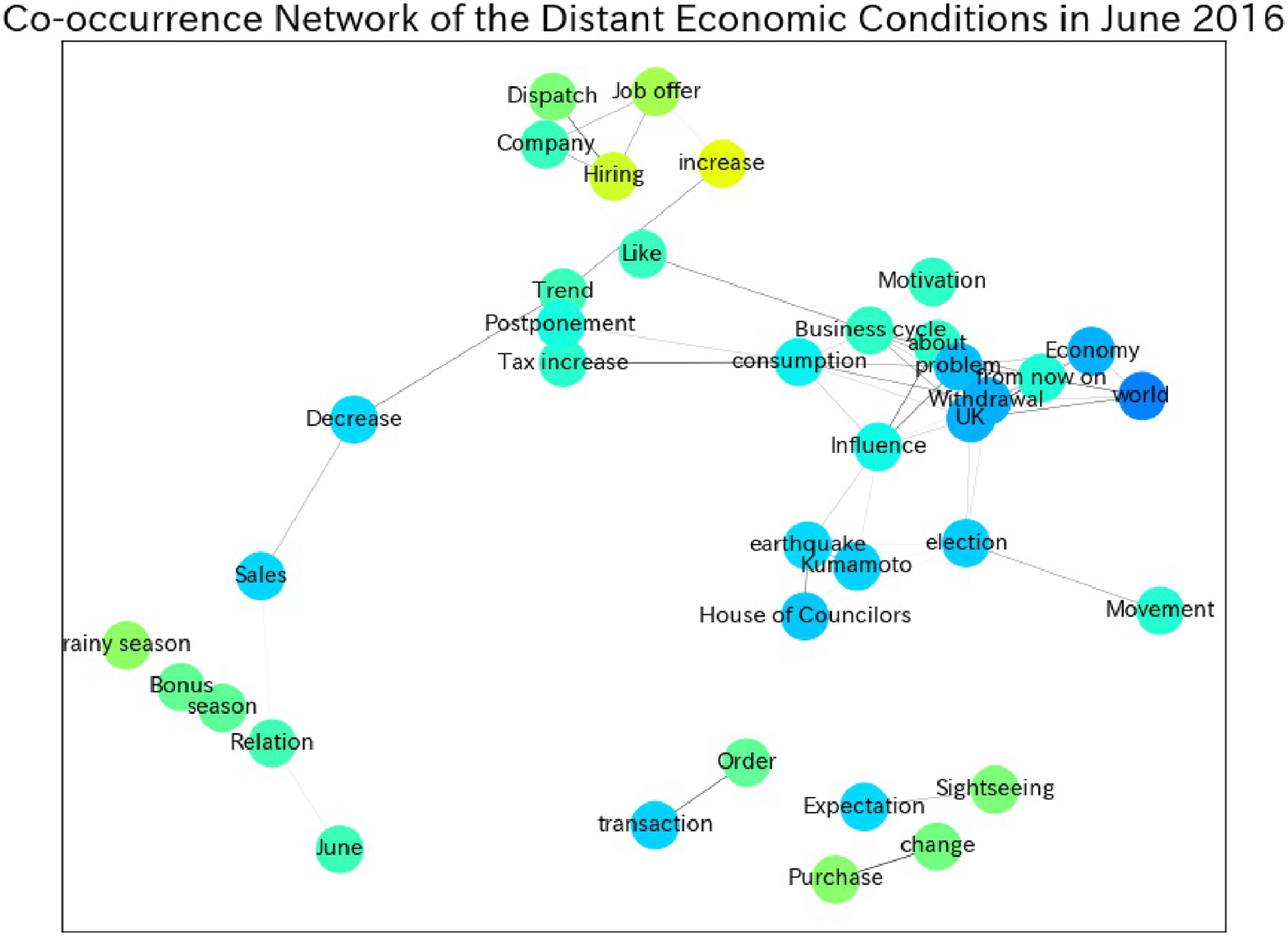}
\includegraphics[width=115mm]{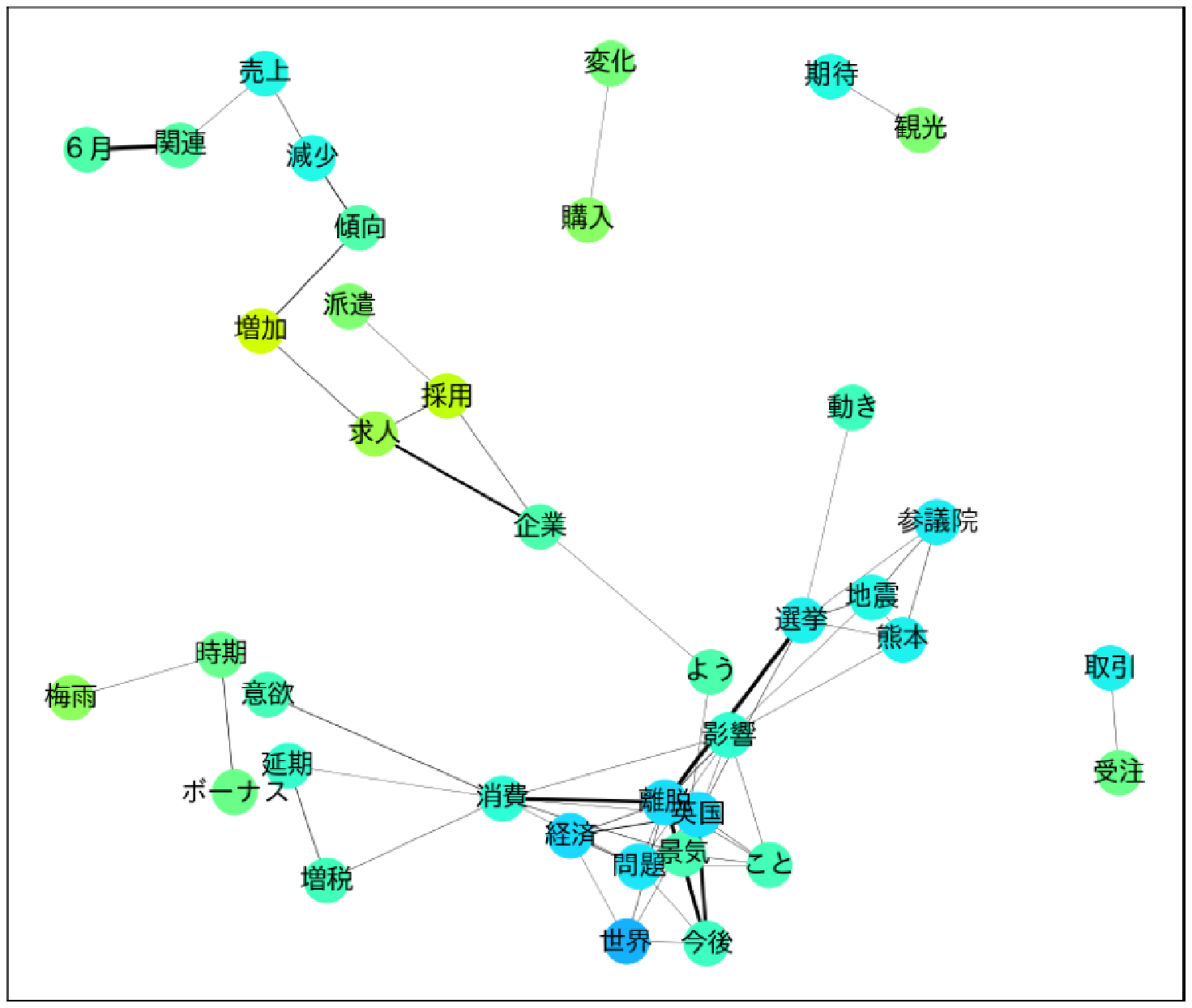}
 \end{center}
\end{figure}
\clearpage

\subsection{Co-occurrence Network of Near Future Economic Conditions in February 2017}
\begin{figure}[H]
\begin{center}
\includegraphics[width=143mm]{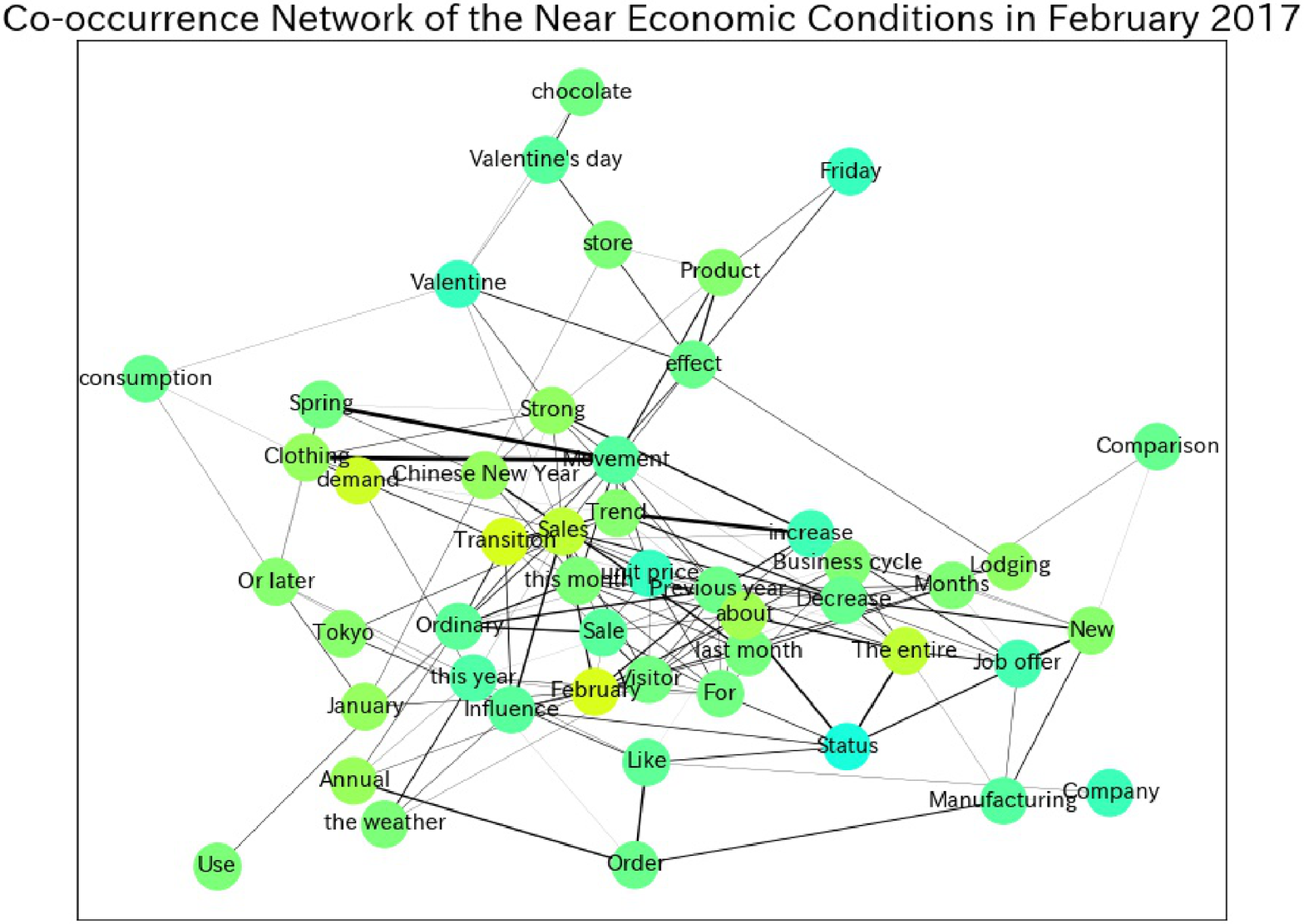}
\includegraphics[width=115mm]{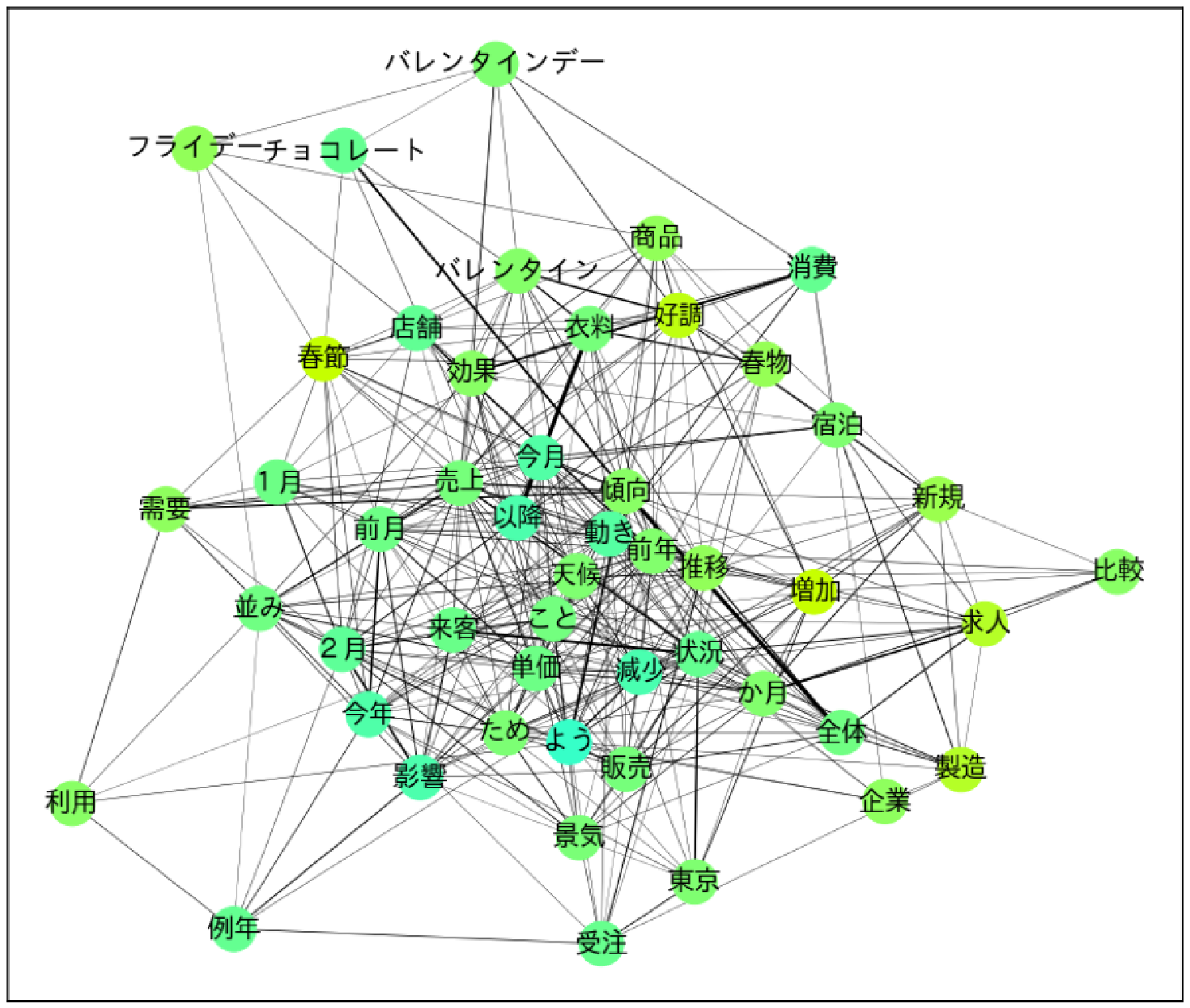}
 \end{center}
\end{figure}
\clearpage

\subsection{Co-occurrence Network of Distant Future Economic Conditions in February 2017}
\begin{figure}[H]
\begin{center}
\includegraphics[width=143mm]{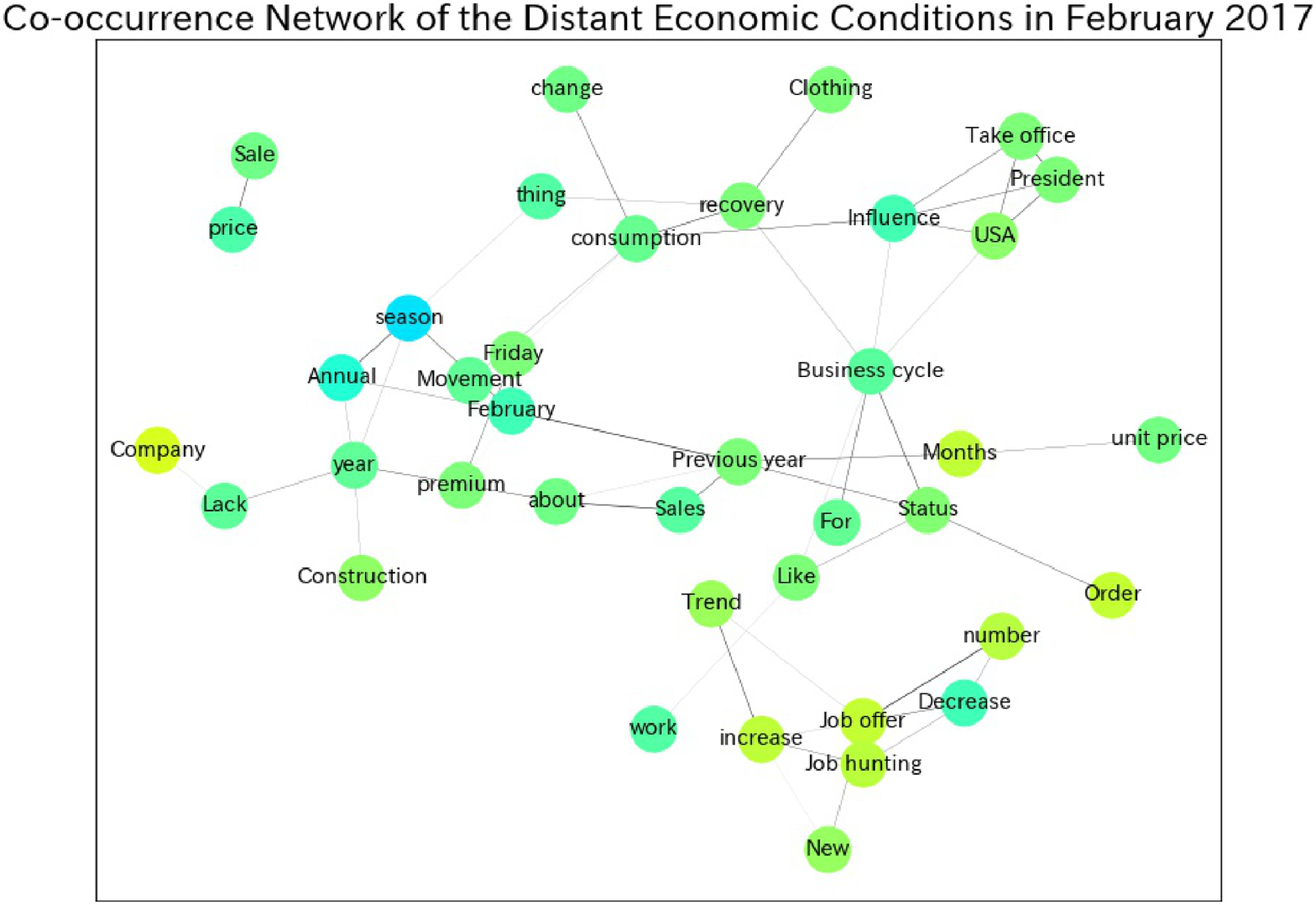}
\includegraphics[width=115mm]{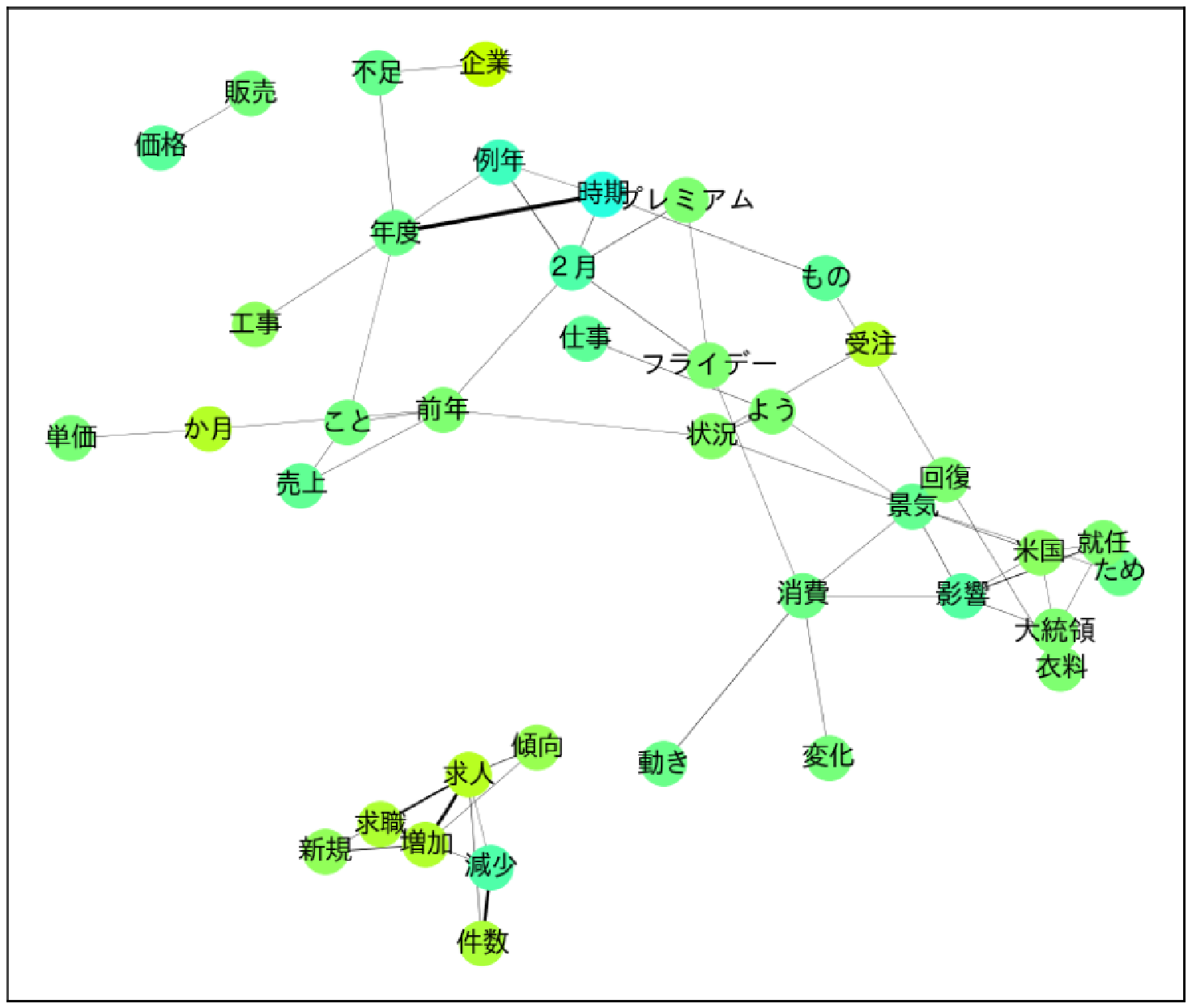}
 \end{center}
\end{figure}

\end{document}